\documentclass[aps,preprint,amsmath,amssymb]{revtex4}
\usepackage{graphicx}
\begin{document}

\title{Analysis of the anomalous electromagnetic moments of the tau
lepton in $\gamma p$ collisions at the LHC}

\author{M. K\"{o}ksal}
\email[]{mkoksal@cumhuriyet.edu.tr} \affiliation{Department of
Optical Engineering, Cumhuriyet University, 58140, Sivas, Turkey}

\author{S. C. \.{I}nan}
\email[]{sceminan@cumhuriyet.edu.tr} \affiliation{Department of
Physics, Cumhuriyet University, 58140, Sivas, Turkey}

\author{A. A. Billur}
\email[]{abillur@cumhuriyet.edu.tr} \affiliation{Department of
Physics, Cumhuriyet University, 58140, Sivas, Turkey}

\author{Y. \"{O}zg\"{u}ven}
\email[]{phyozguvenyucel@gmail.com}
\affiliation{Department of
Physics, Cumhuriyet University, 58140, Sivas, Turkey}

\author{M. K. Bahar}
\email[]{mussiv58@gmail.com} \affiliation{Department of Energy Systems Engineering,
Karamanoglu Mehmetbey University, 70100, Karaman, Turkey}

\begin{abstract}
In this study, we investigate the potential of the process
$pp\rightarrow p\gamma^{*} p\rightarrow p \tau \bar{\nu}_{\tau}q^{\prime} X$ at the LHC  to examine the anomalous electromagnetic moments of the tau
lepton. We obtain $95\%$
confidence level bounds on the anomalous coupling parameters with
various values of the integrated luminosity and center-of-mass energy. The improved bounds have been obtained on the anomalous coupling parameters of electric and magnetic moments of the tau lepton $a_{\tau}$ and $\vert$$d_{\tau}$$\vert$ compared to the current experimental sensitivity bounds. The $\gamma p$ mode of photon reactions at the LHC have shown that it has great potential for the electromagnetic dipole moments studies of the tau lepton.

\end{abstract}

\maketitle

\section{Introduction}

The results obtained in the experimental studies of the anomalous magnetic moments of leptons contain both the estimated values and the new physics contributions, which can not be predicted by the Standard Model (SM). The tau lepton is more advantageous than other leptons in determination of new physics effects since the tau lepton has a larger mass. In many new physics theories, new contributions arising from the anomalous magnetic moment for a lepton with  mass $m$ are proportional to $m^2$. For this reason, since the mass of the tau lepton is much heavier than other leptons, providing anomalous magnetic moment of the the tau lepton to be more sensitive to electroweak and new physics loop contributions.  Additionaly, the tau lepton has a much shorter life time than other leptons, so it is extremely difficult to measure the magnetic moment of the tau lepton by using spin precession experiments. Instead of spin precession experiments, high energy accelerator experiments have been done which include pair production of tau leptons. However, in these experimental studies, $a_{\tau}$ can not be measured directly, since $\tau\bar{\tau}\gamma$ contains off-shell photon or tau leptons (photon virtuality $Q^2=10^5-10^7$ GeV$^2$). The most sensitive experimental bounds have been obtained for $a_{\tau}$ through the process $ e^{+}e^{-}\rightarrow e^{+}e^{-}\tau^{+}\tau^{-} $  at the $95\%$ Confidence Level (C.L.) in LEP is only of $O(10^{-2})$ \cite{L3,opal,del};

\begin{center}
L3: $-0.052<a_{\tau}<0.058$, \\
OPAL: $-0.068<a_{\tau}<0.065$,\\
DELPHI: $-0.052<a_{\tau}<0.013$.
\end{center}

Given these conditions, it can be said that the use of accelerator is more suitable to examine the anomalous magnetic moments of the tau lepton.

The SM theoretical prediction of the anomalous magnetic moment of the tau lepton can be found by summing of all following additives \cite{11,12,hamzeh,samu}:

\begin{eqnarray}
a_{\tau}^{QED}=117324 \times 10^{-8},
\end{eqnarray}
\begin{eqnarray}
a_{\tau}^{EW}=47 \times 10^{-8},
\end{eqnarray}
\begin{eqnarray}
a_{\tau}^{HAD}=350.1 \times 10^{-8}.
\end{eqnarray}.

\noindent Hence, the SM value is obtained as $a_{\tau}^{SM}=0.00117721$. Since this value is far from the experimental sensitivity bounds, which is an order of magnitude below leading QED calculations, more precise experimental measurements should be made.

Another interaction between the tau lepton and photon is CP violating interaction which is induced by the electric dipole moment $|d_{\tau}|$. The SM does not have sufficient information about origin of this interaction \cite{chris}. Since CP violating dipole moment highly suppressed in the SM (induces with three loop level \cite{koba}), any measurement at colliders of the electric dipole moment of the tau lepton gives hints about the new physics beyond the SM. The CP violation may come from new physics theories in lepton sector such as leptoquark \cite{18, 19}, SUSY \cite{20}, left-right symmetric \cite{21, 22} and Higgs models \cite{23, 24,25,26}.

The value of the electric dipole moment of the tau lepton  in the SM is obtained as $|d_{\tau}|\leq10^{-34} e\,cm$ \cite{hoo}.
In addition, the most restrictive experimental sensitivity bounds for $|d_{\tau}|$  have been obtained as follows \cite{belle},
\begin{center}
 $-2.2<Re(d_{\tau})<4.5 \times (10^{-17}\, e\,cm)$,\\
$-2.5<Im(d_{\tau})<0.8 \times (10^{-17}\, e\,cm)$.
\end{center}

\noindent These results have been measured through the process $ e^{+}e^{-}\rightarrow\gamma\rightarrow\tau^{+}\tau^{-} $ by BELLE collaboration \cite{belle}.
Since the value of $Q^2$ in this process is very large ($100$ GeV$^2$), the obtained bounds are given in two parts, as real and virtual.

\begin{figure}
\includegraphics{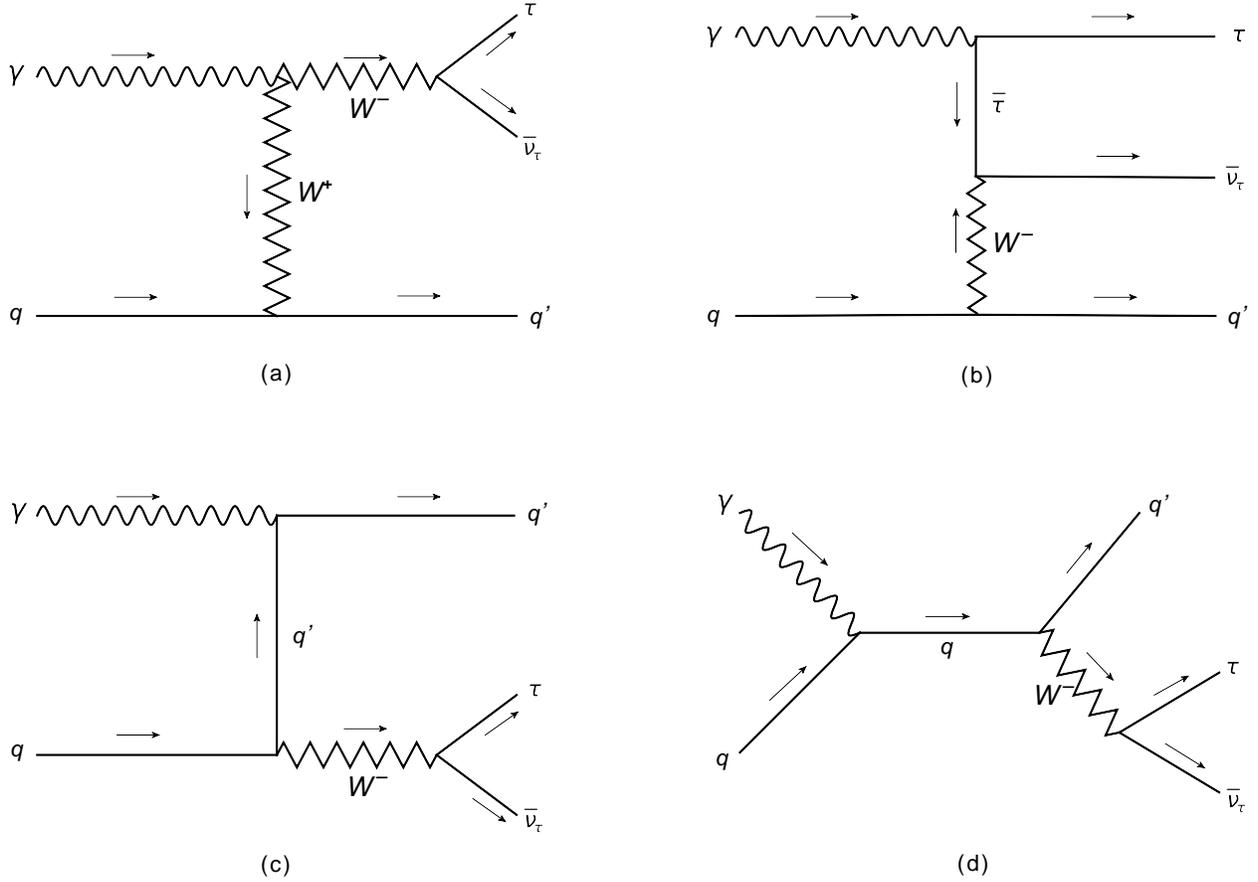}
\caption{Feynman diagrams of the subprocess $\gamma^{*} q\rightarrow  \tau \bar{\nu}_{\tau}q^{\prime} X$ .
\label{feyn}}
\end{figure}

Feynman diagrams of the process $pp\rightarrow p\gamma^{*} p\rightarrow p \tau \bar{\nu}_{\tau}q^{\prime} X$ are shown in Fig.\ref{feyn} and it is clear that the the anomalous electromagnetic
moments contribution of the tau lepton comes from only the diagram (b). The photon in $\bar{\tau}\tau\gamma^*$ vertex in this diagram is Weizsacker-Williams photon. These photons have a very small virtuality ($Q_{max}^{2}=2$ GeV$^2$), as details are explained below.

The most general anomalous vertex function describing $\tau\bar{\tau}\gamma$ interaction for two on-shell
tau and a photon with photon momenta $q$ and mass of the tau lepton $ m_{\tau} $ can be given in the following form \cite{hu,fa},

\begin{eqnarray}
\Gamma^{\nu}=F_{1}(q^{2})\gamma^{\nu}+\frac{i}{2 m_{\tau}}F_{2}(q^{2}) \sigma^{\nu\mu}q_{\mu}+\frac{1}{2 m_{\tau}}F_{3}(q^{2}) \sigma^{\nu\mu}q_{\mu}\gamma^{5},
\end{eqnarray}

\noindent where $\sigma^{\nu\mu}=\frac{i}{2}(\gamma^{\nu}\gamma^{\mu}-\gamma^{\mu}\gamma^{\nu})$ and $ F_{1,2,3}\left( q^{2}\right)$ are the electric charge, the magnetic dipole and electric dipole form factors of the tau lepton, respectively. As known, electromagnetic form factors are given as $ F_{1}=1, F_{2}=F_{3}=0$ in the SM. Usually, form factors are not physical quantities due to the fact that they can contain infrared divergences \cite{mas,bon}. However, when taking into account the limiting case of $ Q^{2}\rightarrow0 $, the form factors become measurable and called dipole moments. These are described through the static properties of the fermions \cite{pich},

\begin{eqnarray}
\label{eqf}
F_{1}(0)=1,\: F_{2}(0)=a_{\tau},\: F_{3}(0)=\frac{2m_{\tau}d_{\tau}}{e}.
\end{eqnarray}

\noindent The kinematical situation (all particles on-shell) relevant to the static dipole moments (\ref{eqf}) can not be realised for the tau lepton at a collider experiment as the mentioned above. To study the signature of the dipole couplings and to compute sensitivity bounds one adopts in a model independent way by means of the effective Lagrangian method. In this study, we will use dimension-six operators specified in Ref.\cite{grz} for the the electromagnetic moments of the tau lepton. Only two of these operators used in Ref.\cite{grz} directly contribute to the electromagnetic dipol moments of the tau lepton at tree level:

\begin{eqnarray}
Q_{LW}^{33}=(\bar{\ell_{\tau}}\sigma^{\mu\nu}\tau_{R})\sigma^{I}\varphi W_{\mu\nu}^{I}
\end{eqnarray}

\begin{eqnarray}
Q_{LB}^{33}=(\bar{\ell_{\tau}}\sigma^{\mu\nu}\tau_{R})\varphi B_{\mu\nu}.
\end{eqnarray}

\noindent where $\varphi$ and $\ell_{\tau}$ are the Higgs and the left-handed $SU(2)$ doublets, $\sigma^{I}$ are the Pauli
matrices and  $W_{\mu\nu}^{I}$ and $B_{\mu\nu} $ are the gauge field strength tensors. Thus, the effective Lagrangian is parameterized as follows,

\begin{eqnarray}
L_{eff}=\frac{1}{\Lambda^{2}} [C_{LW}^{33} Q_{LW}^{33}+C_{LB}^{33} Q_{LB}^{33}+h.c.].
\end{eqnarray}

\noindent After the electroweak symmetry breaking, contributions to the anomalous magnetic and electric dipole moments of the tau lepton are given by

\begin{eqnarray}
\kappa=\frac{2 m_{\tau}}{e} \frac{\sqrt{2}\upsilon}{\Lambda^{2}} Re[\cos\theta _{W} C_{LB}^{33}- \sin\theta _{W} C_{LW}^{33}]
\end{eqnarray}

\begin{eqnarray}
\tilde{\kappa}=\frac{2 m_{\tau}}{e}\frac{\sqrt{2}\upsilon}{\Lambda^{2}} Im[\cos\theta _{W} C_{LB}^{33}- \sin\theta _{W} C_{LW}^{33}],
\end{eqnarray}

\noindent where $\upsilon=246$ GeV and $\sin\theta _{W}$ is the weak mixing angle. The CP even parameter $\kappa$ and CP odd parameter $\tilde{\kappa}$ are related to the anomalous dipole moments of the tau lepton via the following relations: 

\begin{eqnarray}
\kappa=a_\tau, \;\;\;\; \tilde{\kappa}=\frac{2m_\tau}{e}d_\tau.
\end{eqnarray}

Proton-proton collisions at the LHC
reach very high luminosity and center-of-mass energy. On the other hand, these collisions have not very clean environment due to the remnants
of both proton beams after the collision.  The resulting jets from these remnants generate certain uncertainties and
make it difficult to realize the signals which may originate from the new physics beyond the SM. On the other hand, photon emitting protons in photon-induced processes, that is to say $\gamma^{*} \gamma^{*}$ and $\gamma^{*} p$, survive intact without decompose into partons. For this reason, the cleanest channel between $\gamma^{*} \gamma^{*}$, $\gamma^{*} p$ and $pp$ processes is $\gamma^{*} \gamma^{*}$. In $\gamma^{*} p$ process, only one of the incoming protons decomposes into partons but other proton survives intact. As a result, since photon-induced processes have better known initial conditions and much simpler final states, they can compensate the advantages of $pp$ process.

$\gamma^{*} \gamma^{*}$ process is generally electromagnetic in nature and this process has less backgrounds with respect to $\gamma^{*} p$ process. However, $\gamma^{*} p$ process can reach much higher energy and effective luminosity compared to $\gamma^{*} \gamma^{*}$ process. This situation may be important for new physics because of the high energy
dependence of the cross section including anomalous couplings. Thus, $\gamma^{*} p$ process is expected to
have a high sensitivity to the anomalous couplings since it has a higher energy reach than $\gamma^{*} \gamma^{*}$ process.

Photons emitted from one of the proton beams in $\gamma^{*} \gamma^{*}$ and $\gamma^{*} p$ processes can be identified in the framework of
the Weizsacker-Williams Approximation (WWA) \cite{ep,budnev,baur}.  Virtuality of the almost-real photons in the WWA is very low ($Q_{max}^{2}=2$ GeV$^2$).
Since protons emit almost-real photons, they do not decompose into partons. In the WWA, almost-real photons have a small transverse momentum.
For this reason, almost-real photons emitting intact protons deviate slightly from the proton beam path.
Photons emitted with very small angles escape without being identified by the central detectors. Hence, in addition to ATLAS
and CMS central detectors, forward detector equipment is needed to detect intact protons. These equipments can detect intact scattered protons with a very large
pseudorapidity. They are planned to be placed $220$ to $440$ meters away from the central detectors in order to detect intact protons in the interval $\xi_{min}<\xi<\xi_{max}$. This interval is known as the acceptance of the forward detectors \cite{albrow, avati}. The new detectors can detect intact scattered
protons with $9.5<\eta<13$ in a continuous range of $\xi$ where
$\xi$ is the proton momentum fraction loss described by
$\xi=(|\vec{p}|-|\vec{p}'|)/|\vec{p}|$; $\vec{p}$ and $\vec{p}'$ are
the momentum of incoming proton and the momentum of intact proton,
respectively. Thus, the energy of the photons that are interacting can be determined. The relation between the transverse momentum and
pseudorapidity of intact proton is as follows,

\begin{eqnarray}
p_{T}=\frac{\sqrt{E_{p}^{2}(1-\xi)^{2}-m_{p}^{2}}}{\cosh \eta}
\end{eqnarray}

\noindent where $m_{p}$ is the mass of proton and $E_{p}$ is the energy of
proton.
Photon-induced processes were investigated experimentally through the
processes $p\bar{p}\rightarrow p\gamma^{*} \gamma^{*} \bar{p}\rightarrow p
e^{+} e^{-} \bar{p}$, $p\bar{p}\rightarrow p\gamma^{*}
\gamma^{*} \bar{p}\rightarrow p \mu^{+} \mu^{-} \bar{p}$,
$p\bar{p}\rightarrow p\gamma^{*} \bar{p}\rightarrow p W W \bar{p}$ and $p\bar{p}\rightarrow p\gamma \bar{p}\rightarrow p
J/\psi (\psi(2S)) \bar{p}$  by the CDF and D0 collaborations
at the Fermilab Tevatron \cite{cdf1,cdf2,cdf3,cdf4,cdf5}. After these studies, the LHC as a $\gamma^{*} \gamma^{*}$ and
$\gamma^{*} p$ colliders has begun to examine the new physics beyond and within the SM.
Hence, the processes $pp\rightarrow p\gamma^{*} \gamma^{*}
p\rightarrow p e^{+} e^{-} p$, $pp\rightarrow p\gamma^{*} \gamma^{*}
p\rightarrow p \mu^{+} \mu^{-} p$, and $pp\rightarrow p\gamma^{*} \gamma^{*}
p\rightarrow p W^{+} W^{-} p$ have been observed at the LHC by the CMS and ATLAS collaboration \cite{ch1,ch2,ch3,ch4,ch5}.
However, new physics researches beyond the SM through $\gamma^{*} \gamma^{*}$ and
$\gamma^{*} p$ processes at the LHC have been analyzed in the literature \cite{lhc1,lhc1a,lhc2a,lhc4,lhc5,lhc7,inanc,inan,bil,bilx,
bil2,kok,inan2,gru,inanc2,ban,epl,inanc3,bil4,inanc4,koks,
hao1,hao2,kok2,ha1,ha2,koka,tas,ins,ph3,ph4}.
There are also a lot of phenomenological studies about the anomalous magnetic moments of the tau lepton \cite{phe1,phe2,phe3,phe4,phe6,phe7,phe8,diaz}.

Our main motivation in this study is to determine the sensitivity bounds on the electromagnetic moments of the tau lepton at the LHC through the process  $pp\rightarrow p\gamma^{*} p\rightarrow p \tau \bar{\nu}_{\tau}q^{\prime} X$. We consider that this process is much more important in the measurement of electromagnetic moment of the tau lepton since photon-induced processes have better known initial conditions and have clean final states.

\section{Cross Sections and Sensitivity Analysis}

In this work, we have investigated anomalous electromagnetic dipole moments of the tau lepton via the process $pp\rightarrow p\gamma^{*} p\rightarrow p \tau \bar{\nu}_{\tau}q^{\prime} X$. In calculations, we have taken into account subprocess $ \gamma^{*}q \rightarrow  \tau \bar{\nu}_{\tau}q^{\prime} X$ ($ q,q'=u,\bar{u},d,\bar{d},s,\bar{s},c,\bar{c}$). The $b$ quark's distribution is not included in the calculations because it's contribution is too small. The anomalous electromagnetic moments contribution of the tau lepton comes from one diagram (diagram b), which is shown in Fig 1. Here, we have used the following kinematic cuts on final state particles,
\begin{eqnarray}
p_{T}^{\bar{\nu\tau}}&&>10 \quad GeV, \nonumber \\
p_{T}^{\tau},p_{T}^{q}&&>20  \quad GeV,  \nonumber \\
|\eta_{\tau},\eta_{q}|&&<2.5.
\end{eqnarray}
\noindent All calculations have performed using the tree level event generator CalcHEP \cite{calchep} by adding the new vertex functions. In addition, we have used CTEQ6L1 \cite{cteq6l1} for the parton distribution functions and the WWA embedded in CalcHEP for the photon spectra. In the numerical calculations, we have taken the input parameters as $M_p=0.938$ GeV, $M_W=80.38$ GeV,  $M_{\tau}=1.777$ GeV. The cross sections as a polynomial in powers of $\kappa$ and $\tilde{\kappa}$ for the two modes $\sqrt{s}=14,33 $ TeV can be given by
\begin{eqnarray}
\sigma_{Tot}( \kappa,\tilde{\kappa})&&=\sigma_2 \kappa^2+ \sigma'_2 \tilde{\kappa}^2+ \sigma_1 \kappa + \sigma_0   
\end{eqnarray}

\noindent where $\sigma^i (\sigma^{'i})$ $i=1,2$  is the anomalous contribution, while $\sigma_0$ is the contribution of the
SM. This provides more precise and convenient information for each process. The BSM cross section must be proportional to $\kappa^2 + \tilde {\kappa^2}$.
For this reason, the $\kappa^2$ and $\tilde {\kappa^2}$  dependence of the BSM cross section can not be distinguished. Hence the coefficients $\sigma_2$ and $\sigma'_2$ should be the same \cite{Fichet}. Numerical computations of the total cross sections versus $\kappa$ and $\tilde{\kappa}$ at $\sqrt{s} =14, 33$ TeV are given in Table I.

\begin{table}[!ht]
\caption{Numerical computations of the total cross sections versus $\kappa$ and $\tilde{\kappa}$ at $\sqrt{s}=$14, 33 TeV.}
\begin{center}
\begin{tabular}{|c| c| c| c| c|| c| c| c| c|}
\hline
Mode               & $\sigma_2$  & $\sigma_1$ & $\sigma_0$  & $\sigma'_2$   \\
\hline
\hline
$\sqrt{s}=14$ TeV  &    $7.00277$       &     $-0.03345$       &      $0.312531$       &  $7.00277$                                  \\
\hline
$\sqrt{s}$=33 TeV  &    $28.5649$     &       $-0.123529$      &     $0.574318$     &    $28.5649$ \\
\hline
\hline\hline
\end{tabular}
\end{center}
\end{table}

In sensitivity analysis, we take into account $\chi^2$ method,

\begin{eqnarray}
\chi^{2}=\left(\frac{\sigma_{SM}-\sigma(\kappa,\tilde{\kappa})}{\sigma_{SM}\delta}\right)^{2},
\end{eqnarray}

\noindent where $\sigma(\kappa,\tilde{\kappa})$ is the total cross section which includes the SM and new physics, $\delta=\sqrt{(\delta_{st})^{2}+(\delta_{sys})^{2}}$; $\delta_{st}=\frac{1}{\sqrt{N_{SM}}}$ is the statistical error and $\delta_{sys}$ is the systematic error. In this work, we have used cumulative distribution function for $ \chi^2$ which has been defined for use with a method of least-squares. We have taken into account Table 39.2 in \cite{pdg} values for coverage probability in the large data sample limit for one and two free parameters.  

Systematic errors are also taken into account in this work. One of the reasons for these errors comes from the identification of the tau lepton in the experiments. As known, there are many decay channels of the tau lepton. The tau lepton decays have more than one particle in the final state. For this reason, this is called tau jets. These decay channels, called one prong and three prong, are divided into two according to the number of charged particles in the final state. These final states include QCD or hadronic backgrounds. The determination of these situations is much more difficult than in the leptonic final states. In other words, it is difficult to identify the tau lepton. Due to these difficulties and complicated background, the tau identification efficiencies are always determined for specific process, luminosity, and kinematic parameters. However, the hadronic decay of the tau jets can be distinguished from other hadronic decays due to their different final state topology.  Tau identifications have been studied at the LHC \cite{atlastau1, atlastau2, cmstau} and International Linear Detector (ILD) \cite{ild}. As mentioned above, these calculations are made for specific processes. Hence, the general values of the kinematic parameters of the detectors have been taken so that the tau lepton can be identified.

Other reasons for these errors are the experimental errors. However, a systematic error for the processes studied in this article has not yet been studied in the LHC. But, the processes $p p \rightarrow pp\mu^{+}\mu^{-} $  for the $\sqrt{s}=13$ TeV have been examined  at the LHC \cite{ch5}. In this article, systematic error has been found around $3\%$. In addition, a systematic error in a phenomenological study was taken as $2\%$ via the process $p p \rightarrow p\gamma^*\gamma^*p \rightarrow p p\tau^{+}\tau^{-}$ \cite{3}. The systematic uncertainty that arises on the signal is $4.8\%$ in this article by summing quadratically all uncorrelated contributions.
Experiments in which DELPHI collaboration have performed on anomalous magnetic and electric dipole moments of the tau lepton have also been based on systematic errors through the process $e^{+}e^{-} \rightarrow
e^{+}e^{-}\tau^{+}\tau^{-}$ \cite{del}. The systematic errors obtained in these experiments for center-of-mass energies between $183$ and $208$ GeV energies are given in the Table II. In the LEP experiment with the L3 detector, the total systematic error was obtained $7\%$
and $9\%$ at center-of-mass energies $161$ GeV$\leqslant \sqrt{s} \leqslant 209$ GeV through the process $e^{+}e^{-}
\rightarrow e^{+}e^{-}\tau^{+}\tau^{-}$.

Finally, such error can result from theoretical uncertainties. These errors come from QED, electroweak and hadronic loop contributions are extremely small ($ \delta_{theoretical}=5.10^{-8}$) \cite{eidel, Roberts:2010zz, Passera:2004bj,Passera:2006gc, fa}. In the light of these discussions, three different systematic error values have been taken into account in our calculations ($\delta_{sys}=3\%,5\%,7\%$). The limits obtained by this work are given in Tables III-IV with these systematic errors.

\section{Conclusions}

The photon induced reactions at the LHC provide us new opportunities to investigate high energy and high luminosity $\gamma^{*} \gamma^{*}$ and $\gamma^{*} p$ interactions at  higher energy than that at any existing collider. These interactions yield fewer backgrounds than $pp$ deep inelastic scattering. With this clean environment, any discrepant signal with the prospect of the SM would be a conclusive clue for new physics beyond the SM.

In this paper, we have searched the tau lepton anomalous dipole moments in a model independent way through the process $pp\rightarrow p\gamma^{*} p\rightarrow p \tau \bar{\nu}_{\tau}q^{\prime} X$ (where $q,q'=u,d,s,c,\bar{u},\bar{d},\bar{s},\bar{c}$) at the LHC. As can be seen from Fig.2 and Fig.3, the total cross sections of the examined processes increase when the center-of-mass enegy increases. The anomalous magnetic dipole moment is asymmetric, and electric dipole moment is symmetric in the cross sections. This situation can be seen from Eq.14. In Figs.4-5, we show contour diagrams for the anomalous $\kappa$ and $\tilde{\kappa}$ couplings.
 It is understood from Tables III-IV that we improve  bounds values with the increasing energy and luminosity values. An interesting point can be analyzed from these tables that the bounds with increasing systematic error values are almost unchanged according to the luminosity values and for the center-of-mass energy values. The reason of this situation is the statistical error which is much smaller than the systematic error for these systematic error values.

However, our bounds on the anomalous magnetic dipole moments are better than the current experimental bounds for systematic uncertainty is $ 0\%$ and in the same order of magnitude for $7 \%$. Therewithal, our best results are close to other studies in the literature \cite{bil,bilx}. For the anomalous electric dipole moment, the best bounds are the same order of magnitude with the experimental bounds.

\section{Acknowledgements}
This work has been supported by the Scientific and Technological Research Council of Turkey (TUBITAK) in the framework of Project No. 115F136.

 \pagebreak

\begin{table}
\caption{Systematic errors given by the DELPHI collaboration \cite{del}.
\label{tabex}}
\begin{ruledtabular}
\begin{tabular}{ccccc}
 & $1997$& $1998$& $1999$& $2000$ \\
\hline
Trigger efficiency& $7.0$& $2.7$& $3.6$& $4.5$ \\
Selection efficiency& $5.1$& $3.2$& $3.0$& $3.0$  \\
Background& $1.7$& $0.9$& $0.9$& $0.9$  \\
Luminosity& $0.6$& $0.6$& $0.6$& $0.6$ \\
Total& $8.9$& $4.3$& $4.7$& $5.4$  \\
\end{tabular}
\end{ruledtabular}
\end{table}

\begin{table}
\caption{ 95\% C.L. sensitivity bounds of the $a_{\tau}$ couplings
for  various center-of-mass energies and integrated LHC
luminosities.  The bounds are showed with no systematic error ($0\%$) and with systematic errors of $3\%$, $5\%$ $7\%$.\label{tab1}}
\begin{ruledtabular}
\begin{tabular}{cccccc}
 $\sqrt{s}$ (TeV) & Luminosity($fb^{-1}$)& $0\%$ & $3\%$ & $5\%$ & $7\%$  \\
\hline
    &   10 & (-0.0372, 0.0420) &(-0.0529, 0.0577) & (-0.0658, 0.0706) &(-0.0771, 0.0819)  \\
    &  50 & (-0.0242, 0.0290) &(-0.0498, 0.0546) & (-0.0642, 0.0690) &(-0.0762, 0.0809)  \\
14 &  100 & (-0.0200, 0.0248) &(-0.0493, 0.0541) & (-0.0640, 0.0688) &(-0.0760, 0.0808)  \\
    & 200 & (-0.0165, 0.0212) & (-0.0491, 0.0539)& (-0.0639, 0.0687) &(-0.0760, 0.0807)  \\
\hline
    &   100 & (-0.0108, 0.0152) &(-0.0325, 0.0368) & (-0.0424, 0.0467) &(-0.0505, 0.0548)  \\
    &  500 & (-0.0067, 0.0110) &(-0.0323, 0.0366) & (-0.0423, 0.0466) &(-0.0504, 0.0547)  \\
33   & 1000 & (-0.0054, 0.0097) &(-0.0323, 0.0366) & (-0.0423, 0.0466) &(-0.0504, 0.0547)  \\
    & 3000 & (-0.0037, 0.0081) &(-0.0323, 0.0366) & (-0.0423, 0.0466) &(-0.0504, 0.0547)  \\
\end{tabular}
\end{ruledtabular}
\end{table}

\begin{table}
\caption{Same as the Table \ref{tab1} but for the $\vert d_{\tau} \vert$.\label{tab2}}
\begin{ruledtabular}
\begin{tabular}{cccccc}
 $\sqrt{s}$ TeV & Luminosity($fb^{-1}$)& $0\%$ & $3\%$ & $5\%$ & $7\%$  \\
\hline
    &   10 & $ 2.16\times 10^{-16}$ & $ 3.01\times 10^{-16}$ & $ 3.70\times 10^{-16}$ & $  4.32\times 10^{-16}$ \\
    &  50 & $ 1.44\times 10^{-16}$ &$ 2.83\times 10^{-16}$ & $ 3.61\times 10^{-16}$ & $  4.26\times 10^{-16}$ \\
14 & 100 & $ 1.22\times 10^{-16}$ & $ 2.81\times 10^{-16}$ & $ 3.60\times 10^{-16}$ & $  4.25\times 10^{-16}$ \\
    & 200 & $ 1.02\times 10^{-16}$ & $ 2.79\times 10^{-16}$ & $ 3.59\times 10^{-16}$ & $  4.25\times 10^{-16}$ \\
\hline
    &   100 & $ 7.12\times 10^{-17}$ & $ 1.96\times 10^{-16}$ & $ 2.45\times 10^{-16}$ & $  2.90\times 10^{-16}$  \\
    &  500 & $ 4.76\times 10^{-17}$ & $ 1.90\times 10^{-16}$ & $ 2.45\times 10^{-16}$ & $  2.90\times 10^{-16}$  \\
33   & 1000 & $ 4.00\times 10^{-17}$ & $ 1.90\times 10^{-16}$ & $ 2.45\times 10^{-16}$ & $  2.90\times 10^{-16}$  \\
    & 3000 & $ 3.04\times 10^{-17}$ & $ 1.90\times 10^{-16}$ & $ 2.45\times 10^{-16}$ & $  2.90\times 10^{-16}$  \\
\end{tabular}
\end{ruledtabular}
\end{table}

\begin{figure}
\includegraphics{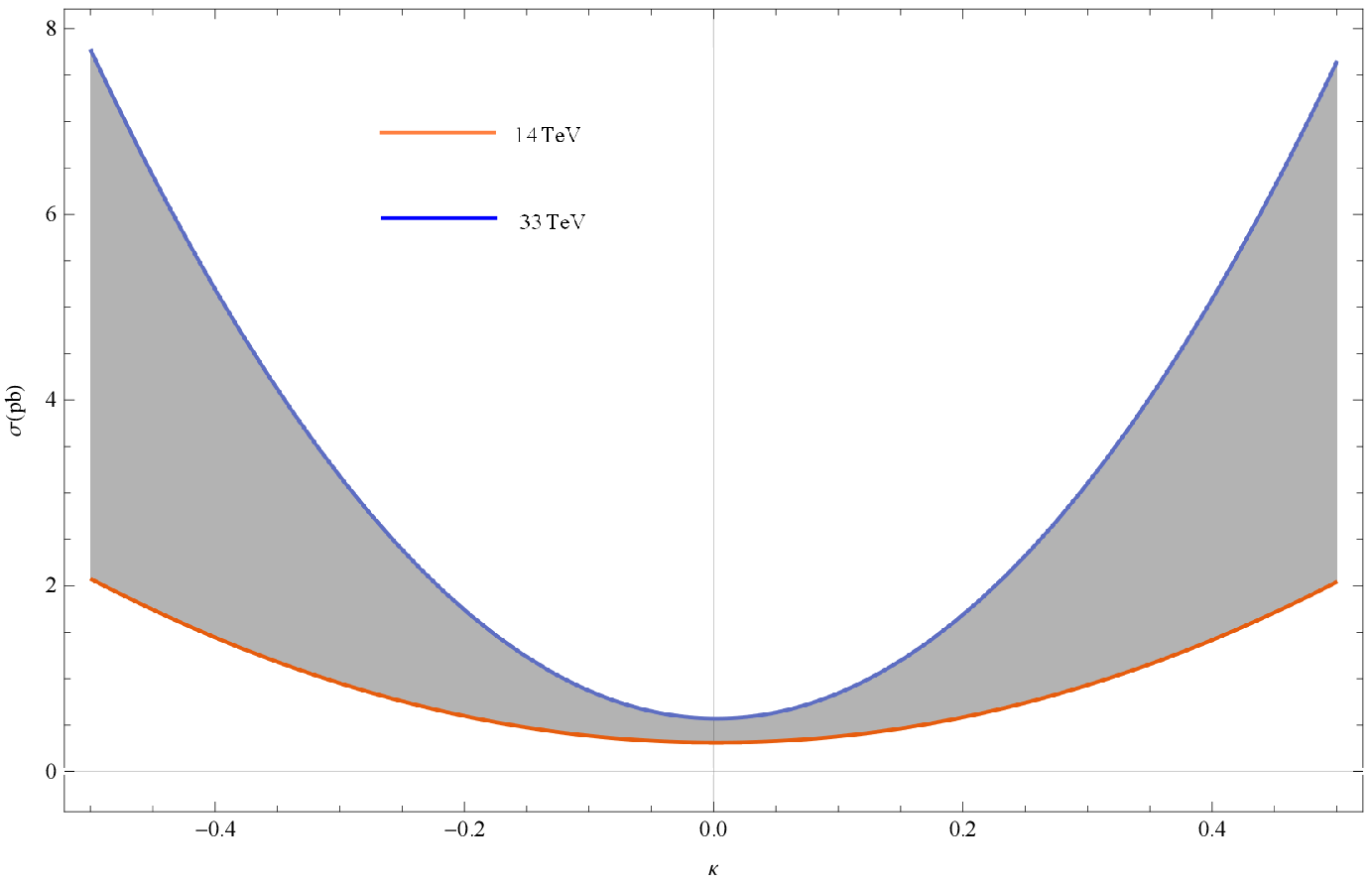}
\caption{The total cross sections of the process
$pp\rightarrow p\gamma^{*} p\rightarrow p \tau \bar{\nu}_{\tau}q^{\prime} X$ as a function of $\kappa$ for two different center-of-mass energies of $\sqrt{s}=14, 33$\hspace{0.8mm}$TeV$.
\label{fig1}}
\end{figure}

\begin{figure}
\includegraphics{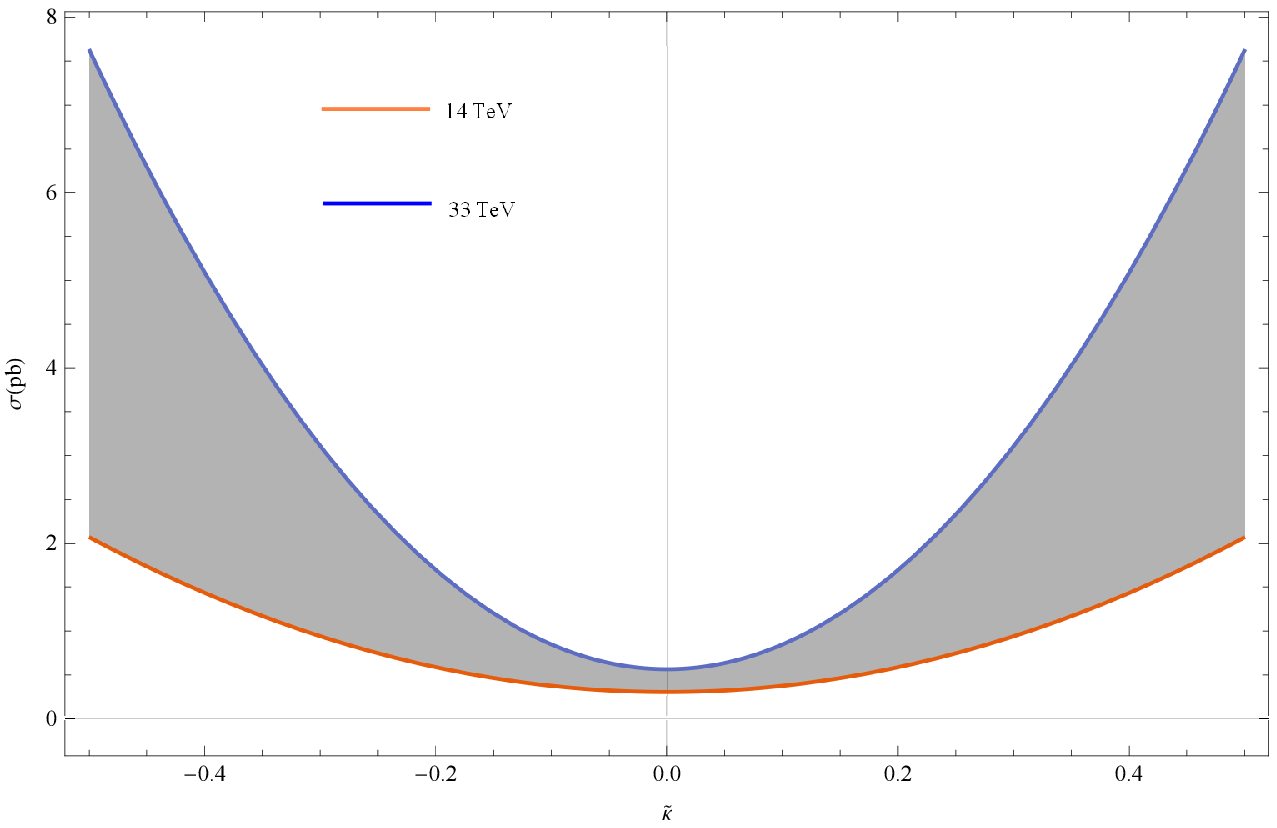}
\caption{The total cross sections of the process
$pp\rightarrow p\gamma^{*} p\rightarrow p \tau \bar{\nu}_{\tau}q^{\prime} X$ as a function of $\tilde{\kappa}$ for two different center-of-mass energies of $\sqrt{s}=14, 33$\hspace{0.8mm}$TeV$.
\label{fig2}}
\end{figure}

\begin{figure}
\includegraphics{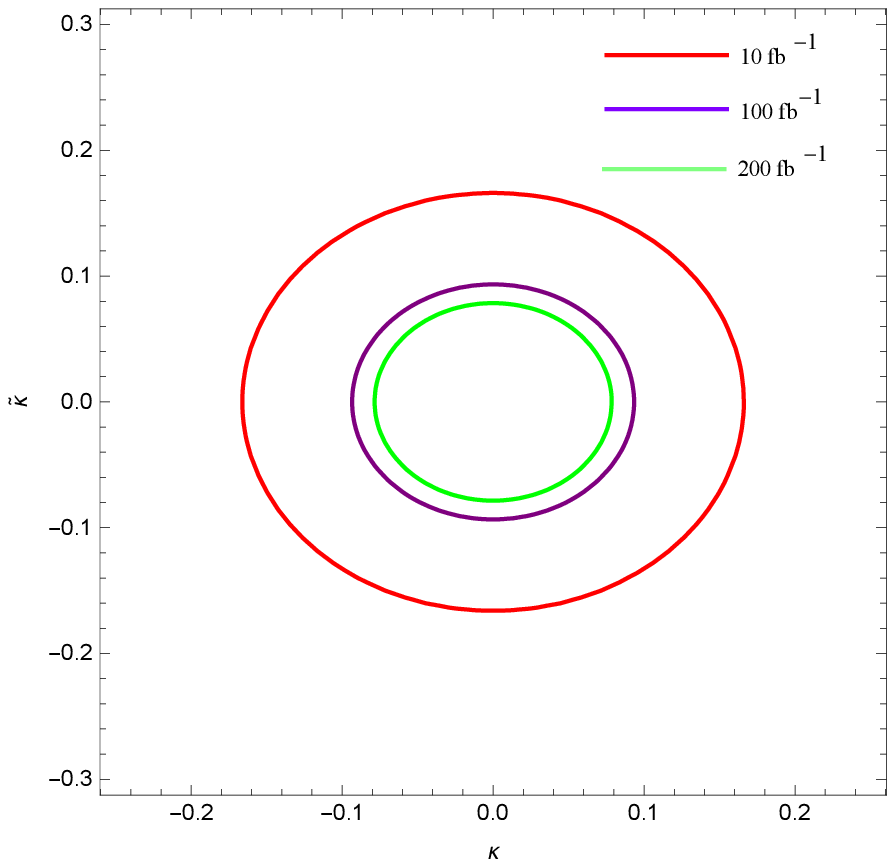}
\caption{Contour limits at the $95 \%$ C.L. in the $\kappa-\tilde{\kappa}$ plane for  $\sqrt{s}=14$ TeV.
\label{fig3}}
\end{figure}

\begin{figure}
\includegraphics{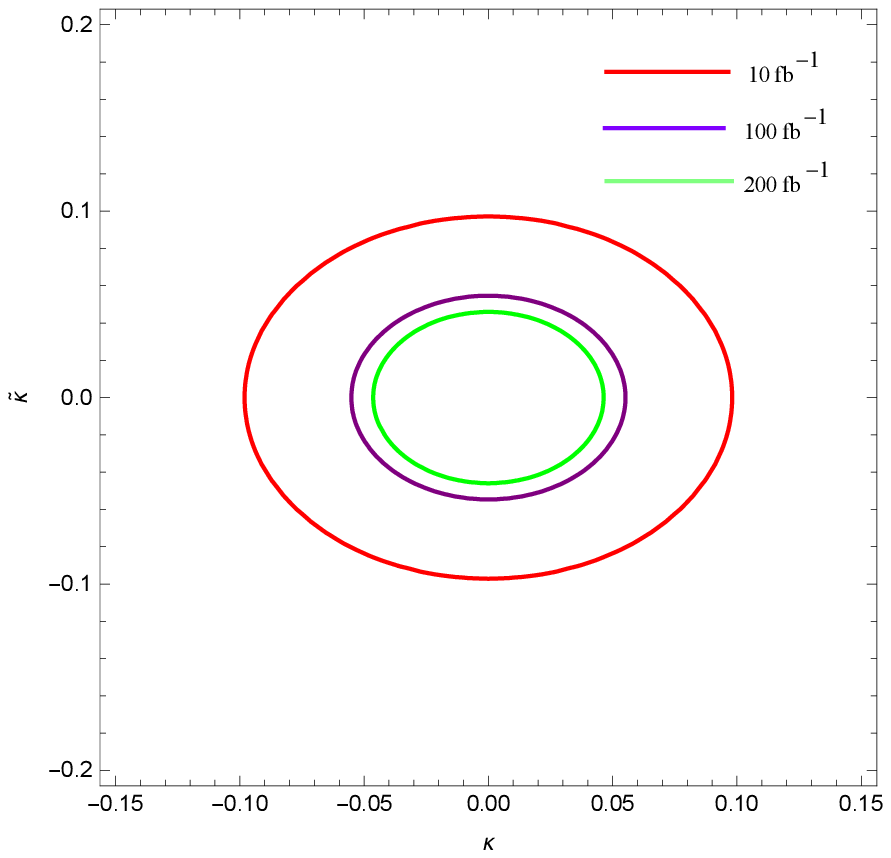}
\caption{Contour limits at the $95 \%$ C.L. in the $\kappa-\tilde{\kappa}$ plane for  $\sqrt{s}=33$ TeV.
\label{fig4}}
\end{figure}

\end{document}